\documentclass[11pt]{article}

\pdfoutput=1

\usepackage{graphicx}

\usepackage[T1]{fontenc}
\usepackage[]{dejavu}

\newcommand{\Darkside}{{DarkSide}} \newcommand{\darkside}{{DarkSide }}
\newcommand{\Fluka}{F{\sc{luka}} } \newcommand{\Flukadot}{F{\sc{luka}}}

\makeatletter
\def\blfootnote{\xdef\@thefnmark{}\@footnotetext}
\makeatother

\begin{document}

 \begin{centering}

 {\LARGE \sffamily   Study of Cosmogenic Neutron Backgrounds at LNGS  \par}

 \vskip 9mm

 {\Large \hskip -\parindent   A. Empl$^{*}$\blfootnote{$^{*}$Corresponding author, Anton.Empl@mail.uh.edu},
        R.  Jasim, E. Hungerford and P. Mosteiro$^{a}$}

 \vskip 3mm
 {\small \hskip -\parindent \hskip 1.7mm  Department of Physics, University of Houston, Houston, TX 77204} \\
 {\small \hskip -\parindent   $^{a}$Department of Physics, Princeton University, Princeton, NJ 08544}

 \end{centering}

 \vskip 15mm
 \hskip -\parindent {\large\bf  Abstract.} \hskip 3mm
 Cosmic muon interactions are important contributors to backgrounds
 in underground detectors when searching for rare events. Typically
 neutrons dominate this background as they are particularly difficult
 to shield and detect in a veto system.  Since actual background
 data is sparse and not well documented, simulation studies must be used to
 design shields and predict background rates. This means that validation of
 any simulation code is necessary to assure reliable results. This work
 studies the validation of the FLUKA simulation code, and
 reports the results of a simulation of cosmogenic background for a liquid
 argon two-phase detector embedded within a water
 tank and liquid scintillator shielding.

 \vskip 20mm


\section{Introduction}

This paper reports the use of the FLUKA simulation code to study muon-produced,
cosmogenic backgrounds at the  Italian National Laboratory for
Underground Research  (LNGS). Care was taken to validate FLUKA with available
data, and compare the results of our studies with previous simulations. We are
particularly interested in the predicted neutron backgrounds for a
dark matter direct detection experiment \Darkside-50, now under construction at the
LNGS, and 
its potential extension to a 3 ton liquid argon two-phase detector. We also
report work relating to backgrounds in the nearby Borexino
detector, which when this data is available in the near future, will
provide substantially better cosmogenic neutron information which can
be used to validate further simulations.

 Cosmic muon interactions can be important contributors to backgrounds
 as they may dominate experiments searching for rare events.   Typically,
 such experiments are placed in deep underground facilities where
 only neutrinos and high energy muons are able to reach the detectors.
 For example, the first measurement of atmospheric neutrinos was 
 undertaken  at the Kolar Gold Fields in 1965 \cite{kolar}.   Muons in
 particular, induce background rates through showers of secondary
 particles created in local interactions near the detector, and only
 high energy muons penetrate to the depth of the experimental
 halls.   Thus the total cosmic muon flux decreases with depth as lower
 energy muons are removed from the spectrum. However, little
 experimental information about muon-induced secondaries  at
 depth is available.

 Most studies have concentrated on the neutron flux in the muon radiation
 field, since energetic neutrons are difficult to shield and are
 usually the critical background.   However, the low rates and
 challenges inherent in neutron measurements require careful
 interpretation of the available data.   All measurements of the
 neutron flux employ liquid scintillation detectors which record
 the gamma radiation emitted after neutron capture. Obviously detector
 geometry and efficiencies affect the interpretation of the data, 
 so experiments have to rely on simulations of
 particle interactions and transport to extract and predict the
 measured muon-induced backgrounds.

\section{FLUKA}

 \subsection{The FLUKA Code}

 \Fluka \cite{fluka1, fluka2} is a fully integrated particle-physics,
 Monte Carlo simulation package which traditionally has been applied
 to problems related to cosmic rays and cosmogenic backgrounds in deep
 underground experiments.   The origin of \Fluka however, is linked to
 shielding designs for particle accelerators, and was started
 as early as 1962.  The modern \Fluka code finds many applications
 ranging from high energy particle physics to medical physics and
 radio-biology.

 Design and development of \Fluka has always focused on the
 implementation and improvement of verified physical models. 
 Thus \Fluka adopts microscopic models connected in a way
 which maintains consistency among all the reaction steps and
 types.   Conservation laws are enforced at each step and predictions
 are bench-marked against experimental data at the level of single
 interactions.   This results in a consistent approach to all
 energy/target/projectile combinations with a minimal      overall
 set of free parameters.   As a consequence, predictions for complex
 simulation problems are robust and arise naturally from the
 underlying physical models, even where no direct experimental data
 are available \cite{manual}.

 Information about the implemented models is available through
 the \Fluka manual, additional documentation, and lectures located
 at the official \Fluka website \cite{fluka0}. The version of \Fluka
 used for the present study is FLUKA2011.2, from November 2011.

 \subsection{Relevant code benchmarks}

 Predictions of backgrounds to rare physics processes at underground facilities
 depend on the full range of physics implemented in \Flukadot.   To
 illustrate the quality of \Fluka predictions, a few relevant published
 benchmark comparisons with available experimental data are 
 presented in the following subsections.

 \subsubsection{Muon-nuclear interactions}

\begin{figure}[bht]
    \centering
    \includegraphics[width=6.5in]{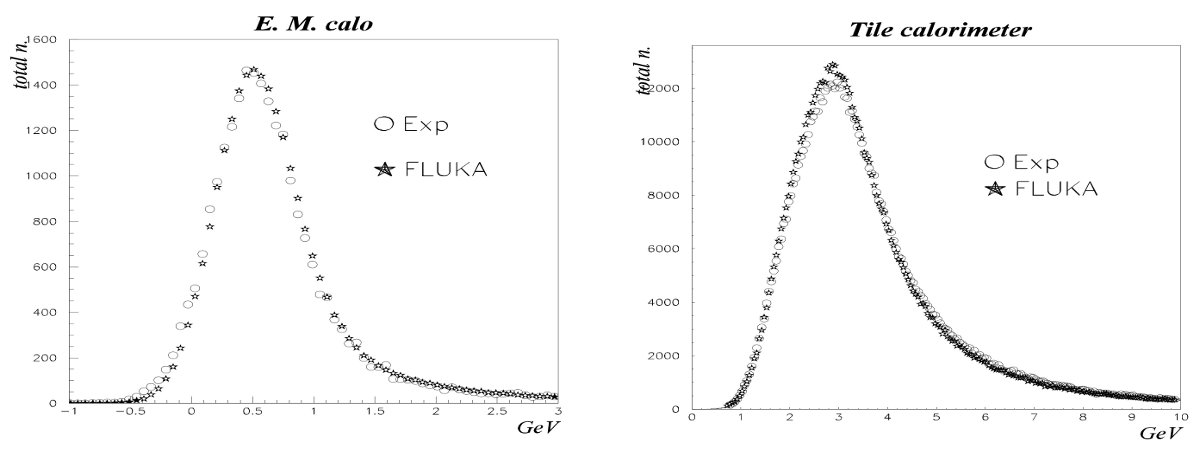}
     \caption{Energy deposition into the electromagnetic (a) and
              hadronic (b) part of the ATLAS
              combined calorimeter from a 300-GeV muon test beam at CERN.}
   \label{fig:atlas}
   \end{figure}

 Although muon-nuclear interactions at high energies are not well
 measured, they are implemented in \Fluka by photo-nuclear interactions
 involving virtual photons. The energies relevant to the simulations
 of interest are approximately 250-300 GeV for single-muon events.
 Figure \ref{fig:atlas} shows the results of the CERN muon test beam of
 300~GeV incident on a section of the ATLAS tile calorimeter~\cite{atlas}.
 The energy deposition,
 resulting from both electromagnetic and hadronic showers,
 is shown in the figure, and agreement between measurement and simulation
 is very good.
 In particular, events with very large energy deposition in the
 hadronic calorimeter are well reproduced.   For this class of events,
 hadronic effects become comparable to electromagnetic effects.   It
 should be noted that this agreement is absolute as no relative scaling
 is applied.    The energy scale for both the measurement
 and the \Fluka simulation are derived independently from calibration
 measurements using a 20-GeV electron beam. 

 \subsubsection{Photo-nuclear interactions}

\begin{figure}[hbt]
    \centering
    \includegraphics[width=3.2in, height=2.7in]{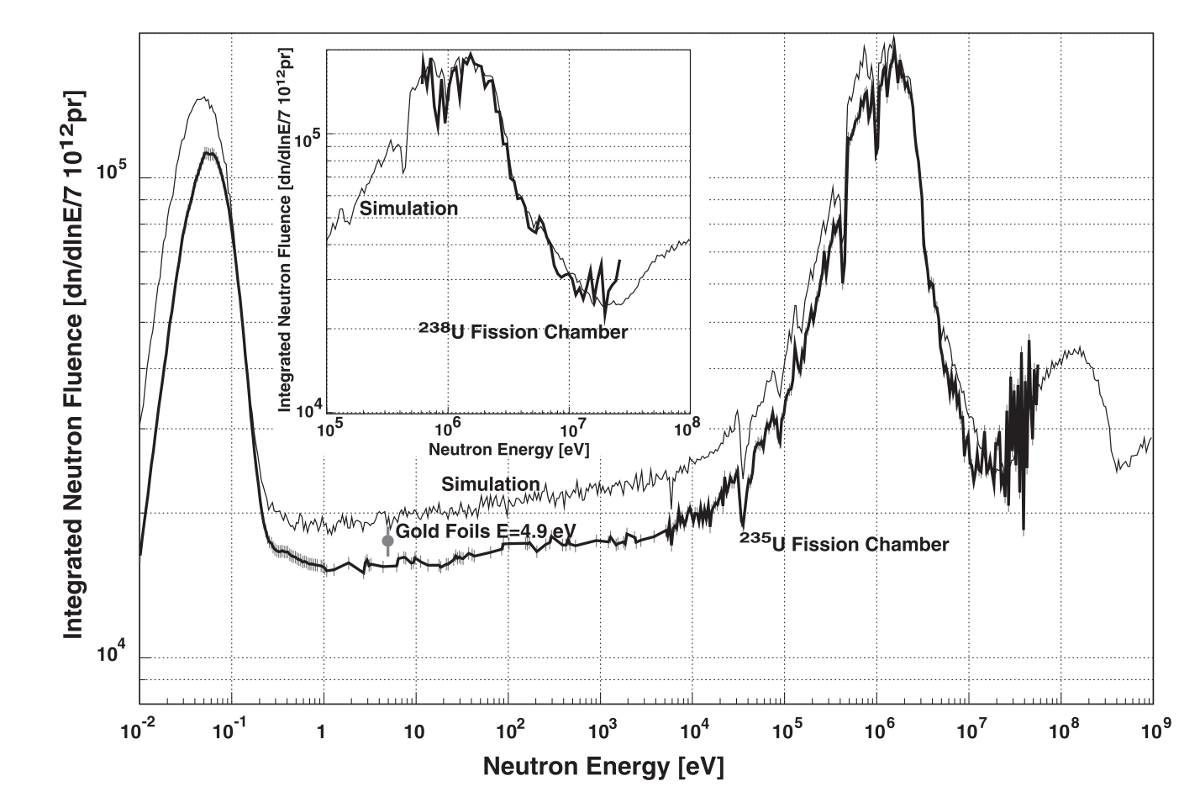}
    \includegraphics[width=3.2in]{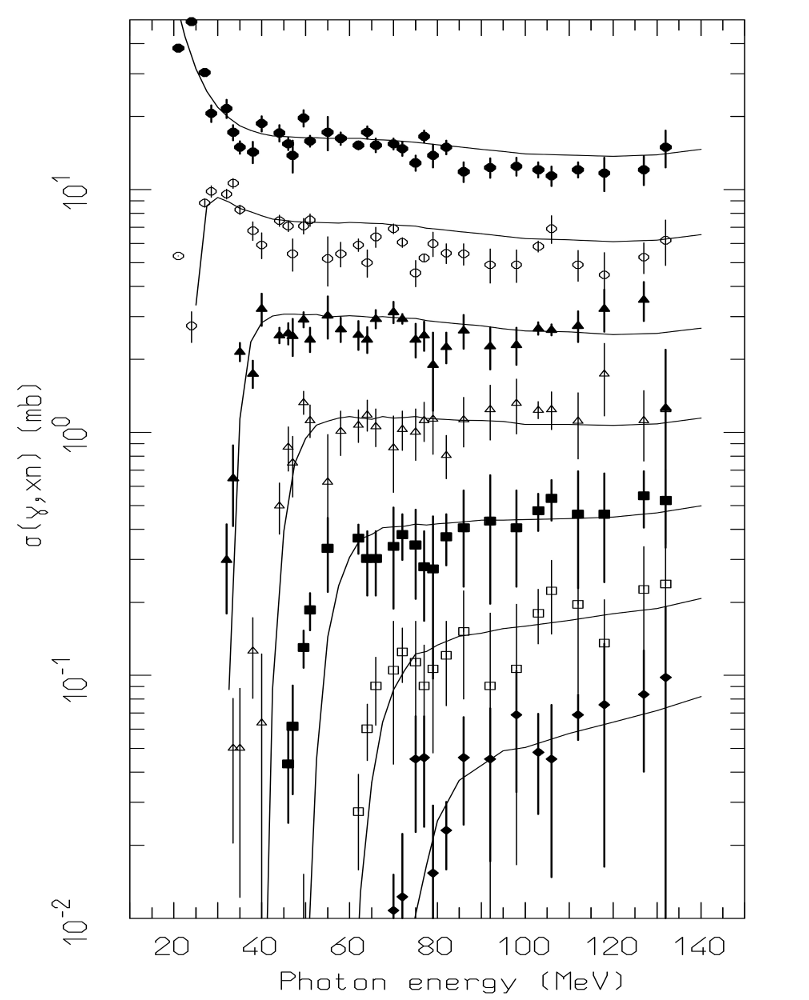}
  \caption{ (a) Neutron fluence as a function of energy from
           measurements with \textsuperscript{235}U
           and \textsuperscript{238}U fission chambers. The dashed
           curve represents the simulated
           fluence. The value of the fluence at 4.9 eV as determined
           by the activation of gold
           foils is also shown. \,\, (b) Excitation functions
           for emission of $N$ or more neutrons
           from a lead target hit by mono-energetic photons.  See text
           for more details. }

   \label{fig:ntof}
   \end{figure}

 Muon-nuclear interactions are treated as virtual photo-nuclear
 interactions in \Flukadot.   Figure \ref{fig:ntof}b shows the results of a
 measurement of photo-nuclear reactions on a lead target,
 \textsuperscript{208}Pb($\gamma$,x n), in the energy range of
 20~MeV~$<E<$~140~MeV.   The cross section for multiple neutron emission as
 a function of photon energy is also shown.   The symbols correspond
 to experimental data, and the \Fluka predictions are lines.   The
 different curves correspond to events with neutron multiplicities
 $\geq N$, with $2\leq N\leq 8$, in descending
 order\footnote{The individual results were scaled by $ 1/(N-1)$ with
 $N$ being the respective neutron multiplicity.}.
 Data and simulation are in good agreement.   See Ref. ~\cite{fluka3}
 for more details on the implementation of photo-nuclear interactions
 in \Flukadot.

 \subsubsection{Neutron production and transport}

 Neutron interaction and transport are     important to obtain
 an accurate simulation of muon-induced
 neutron backgrounds.  The majority of neutrons are produced by secondary
 processes in hadronic showers.   To demonstrate \Flukadot's capabilities for
 simulating neutron production and transport, Figure \ref{fig:ntof}a
 shows the measured and simulated neutron energy spectra for the n\_TOF
 facility\cite{ntof1}.   This neutron time-of-flight facility is a
 high-intensity neutron source which has been operating at CERN since
 2001 \cite{ntof}.   Neutrons are produced by a 20 GeV pulsed proton
 beam interacting with a solid lead target.  The detectors are located
 about 180 meter downstream.   The spectrum predicted by \Fluka is the
 result of a simulation which starts with interacting protons in the lead
 target and includes the entire experimental geometry with all relevant
 physics processes.

 The agreement between simulated and measured spectrum is apparent,
 although a 20\%
 discrepancy was found in the energy range from $\approx$  0.4 eV  to a
 few MeV. Later, it was noted that a water absorber just
 downstream of the neutron production target was 1 cm thicker than
 recorded on the design drawings.  When this error was corrected 
 the initial discrepancy in the simulation was removed.


 \subsection{FLUKA physics options}

 The physics models in \Fluka are fully integrated into the code, so that
 the user is presented with an optimal approach in the opinion of the
 developers.   Several default settings addressing different general physics
 problems are available.   These can be further modified from the default
 behavior via \Fluka options.

 For this study, the simulation was performed with the \Fluka default setting
 PRECISIO(n).   In addition, photo-nuclear interactions were enabled through
 the \Fluka option PHOTONUC and a more detailed treatment of nuclear
 de-excitation was requested with the EVAPORAT(ion) and COALESCE(nce)
 options.   The latter two options are suggested to the user in order to
 obtain more reliable results for isotope production.   These enable the
 evaporation of heavy fragments (A$>$1) and the emission of energetic
 light-fragments, respectively.  The treatment of nucleus-nucleus
 interactions was also turned on for all energies via the option IONTRANS,
 and delayed reactions were enabled through the option, RADDECAY.

 To evaluate the muon-induced neutron production rate in the simulation,
 we record neutron captures on hydrogen inside     liquid
 scintillator.   This approach closely follows that of the experimental 
 measurement and avoids technical ambiguities for neutron counting.


 \subsection{Simulation concerning liquid scintillator}

 As previously noted,     all available experimental information 
 about cosmogenic neutrons and muon-induced secondary products 
 are obtained from experiments through the use of liquid scintillator.
 Depending
 on the respective detector configuration, the surrounding materials may
 also significantly contribute to the flux. With respect to \Flukadot, the
 following two issues are identified.

 \subsubsection{Treatment of deuterons in FLUKA}

 The user can vary the degree of detail for the simulation of nuclear
 processes in \Fluka in order to optimize CPU time constraints and
 the nature of the physics problem.   Because of the focus on
 neutron and isotope production, a realistic treatment of nuclear de-excitation
 is mandatory to study cosmogenic neutron production.   As a result
 \Fluka produces heavy fragments (A\,$>$\,1) in addition to other particles
 in the process of returning nuclei to their ground state.  In particular
 for liquid scintillator, a large fraction of the
 heavy fragments are energetic deuterons.   For the simulation of neutron
 production in liquid scintillator by cosmogenic muons with a mean energy
 of 283~GeV, the number of neutrons in the final state is reduced by
 approximately 8\% when fragments with A\,$>$\,1 are requested.
 However, the current version of \Fluka implements transport but no
 interaction model for deuterons.  Consequently, all neutrons which are
 incorporated into deuterons are lost, though some neutrons should be
 partially recovered through deuteron spallation.

 \subsubsection{Photoproduction of $^{11}$C}

 In \Flukadot, nuclear de-excitation is performed according to a Fermi break-up
 model for light nuclei having mass number A$<$16.   An updated
 version of this model will be available with the next major \Fluka release.
 The updated model implements additional conservation laws as well as
 constraints on available final state configurations and
 symmetries~\cite{sala}.
 As a relevant example, the $ \gamma \, + \, ^{12}C $
 reaction in the current FLUKA model strongly
 favors break-up of $^{12}C$ into 3\,$\alpha$ over neutron emission
 via photo-production.   In this case the parity and spin of
 the 3 $\alpha$ breakup are:
   $$ \gamma \, (1^{-}) + ^{12}C \, (0^{+}) \,
              \rightarrow \,3 \,\alpha (\, 0^{+}). $$
Thus even though this is energetically and model favored, the 3\,$\alpha$
break-up with L=0 is forbidden.  As a result, the reaction \,
$^{12}$C($\gamma,n$)$^{11}$C \, is under-predicted by current and
previous \Fluka versions.


\section{Cosmogenic background simulation at LNGS}

 A faithful simulation of the muon radiation field in the vicinity of
 any underground experiment requires detailed information of the depth,
 overburden geometry, and composition of the surrounding rock through
 which the muons propagate. While the radiation field can be assumed
 constant at the average depth of the experimental hall, since the
 cavern size scales are small relative to changes in the flux, details
 of the detector geometry and materials surrounding the cavern and detector are
 important. 

\subsection{Previous FLUKA Simulations}

 Most all approaches to simulating muon-induced neutron backgrounds in
 deep underground experiments use a general parameterization of the neutron
 radiation field in the cavern, as given, for example, in Ref.~\cite{hime}.
 The neutron radiation field is created by single muons but other
 components of the cosmogenic event, and site specific information
 are not implemented.
 However, maintaining the complete radiation field significantly modifies the
 character of the background problem. This can be seen, for example, in the
 issue involving the $ \gamma \, + \, ^{12}C $ reaction, as pointed out in
 the previous section.

\begin{figure}[htb]
  \centering
  \includegraphics[width=2.5in]{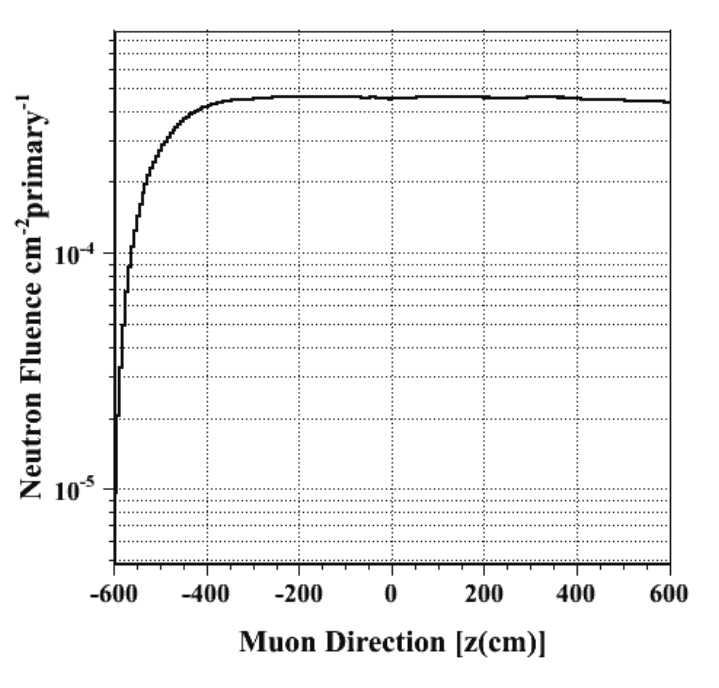}
  \includegraphics[width=3.95in,height=2.4in]{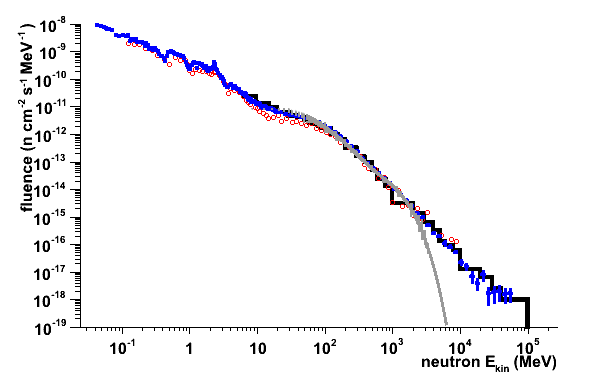}
  \caption{(a) A figure showing the neutron flux as a function of
            the distance traveled by a muon with 270 GeV kinetic
            energy~\protect{\cite{riznia}}.  \,\,
            (b) Neutron kinetic energy spectrum as predicted for a
            cavern at the LNGS laboratory.  Blue symbols with
            statistical errors only: current \Flukadot,
            red symbols: Wulandari et al.~\protect{\cite{wulandari}}, \,
            solid black line: Dementyev et al.~\protect{\cite{dementyev}} \,
            and gray line:~\protect{\cite{hime}}.
          }
  \label{fig:nEkin}
  \end{figure}

 In an initial study of secondaries induced by muons, we used \Fluka to
 reproduce the muon-induced neutron kinetic energy spectrum for LNGS
 following the prescriptions given by
 Ref.~\cite{hime}.   The neutron flux is found for a standardized
 cubical cavern of dimensions 6 m, permitting
 single muons to interact in the surrounding 7-m thick layer of rock.
 The spectra are normalized  to the LNGS muon flux of 1.17~$\mu$/hr/m$^{-2}$
 in  accordance with measurements. A study of the rock thickness
 required to equilibrate neutron production is summarized in
 Figure~\ref{fig:nEkin}a.

 Our simulated neutron spectrum is shown in Figure~\ref{fig:nEkin}b by
 the solid blue symbols 
 along with the parameterization given by
 Ref.~\cite{hime}, indicated by the gray line.  Two more
 predictions from     literature are included in this comparison.  The open
 red symbols present a simulation result prepared in the context of
 the CRESST experiment, also making use of
 \Flukadot~\cite{wulandari}, while the black histogram (for E$_{kin}$\,$>$ 6 MeV)
 was derived by the LVD collaboration, in part using the hadronic transport
 code SHIELD~\cite{dementyev}.   Both experiments were situated at the
 LNGS laboratory.

 Our standard \Fluka results and the predictions for the two experiments are
 in good agreement over the entire available energy range.
 In contrast, the distribution and
 parameterization suggested in Ref.~\cite{hime} do not reproduce the
 behavior predicted by FLUKA at high energies.  Even though there are few
 neutrons at these energies, they are of special concern since these
 energetic neutrons  are very penetrating.   Further note that all results
 in Ref. \cite{hime} include
 an independent increase of the neutron multiplicity beyond FLUKA
 predictions. This correction introduces an extra 30\% to the neutron flux
 and was implemented in Ref.~\cite{hime} to fit the
 observed data.   However, Figure~\ref{fig:nEkin}b  indicates
 that other \Flukadot-based work does not require this additional
 increase to reproduce the neutron kinetic energy spectrum.  

 In general, the previous \Fluka simulations of deep underground
 cosmogenic backgrounds do not address several issues to be discussed
 later, since site and experiment specific information were not included.   

 \subsection{Muon radiation field}

\subsubsection{Particle components of the muon radiation field}

\begin{figure}[htb]
  \centering
  \includegraphics[width=3.30in]{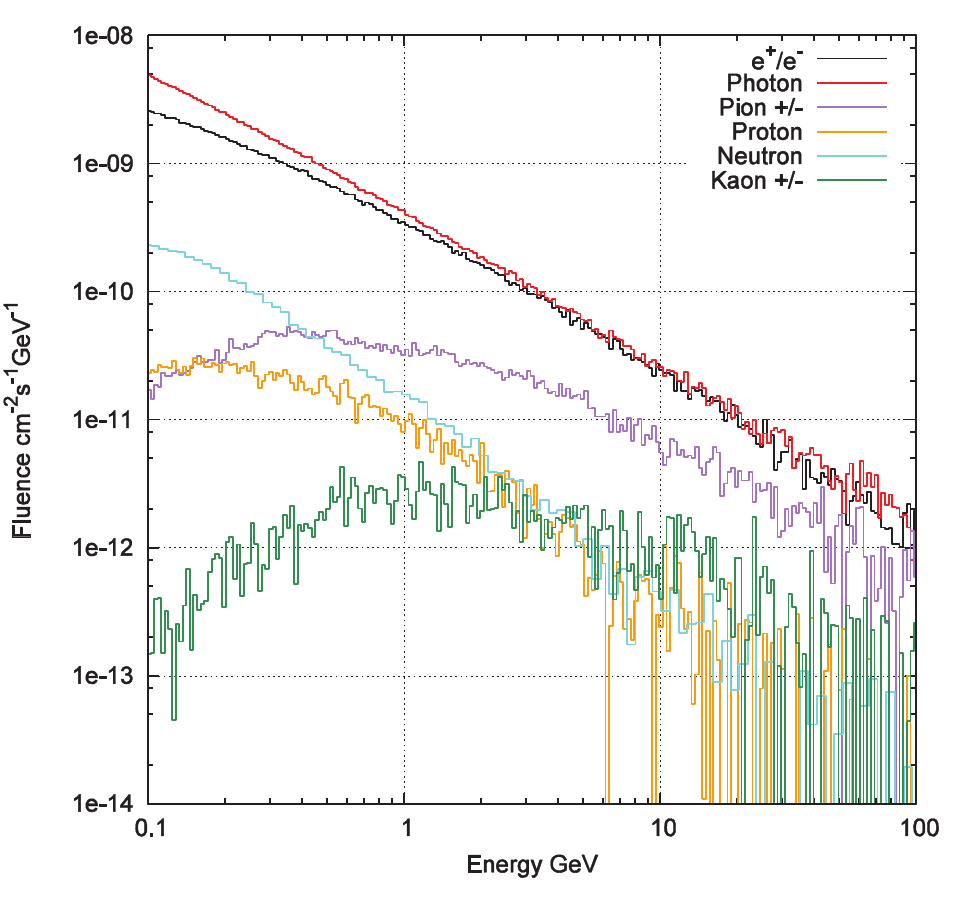}
    \caption{Particle fluence produced by 250 GeV muons on rock}
  \label{fig:particles}
  \end{figure}

As previously indicated, high energy muons produce many particle
types in addition to neutrons. These particles including the primary
muon can continue to produce backgrounds as they interact with the
detector and its surroundings.  Figure~\ref{fig:particles}
shows the particle fluence produced by 250 GeV muons.  Aside from
neutrons, photo-production
can also contribute to backgrounds, including neutron backgrounds due
to the large photon flux even though cross sections are due to
electromagnetic interactions.

   \subsubsection{Intensity}
 
 The muon flux in the 3 halls at LNGS is summarized in a recent
 publication~\cite{seasonal}.   The total muon flux for the simulation
 described in this paper  is normalized to an experimental measurement.
 More precisely, this measured flux describes the total cosmogenic
 muon event rate, since individual counts may contain more than one
 muon~\cite{decoherence}.   Differences in muon flux between these
 measurements as given in Table~\ref{mu_flux} may be attributed to the
 relative location of the halls within the laboratory, but also could
\begin{table}[htb]
 \begin{center}
\caption{\label{mu_flux} The table shows the measured muon flux in the
                         various experimental halls at the LNGS}
   \begin{tabular}{lccl}
    experiment & Hall & year published & total muon event rate \\
               &      &                & ($\times 10^{-4} \,\, s^{-1}
               m^{-2}$) \\ 
    \hline
    LVD      & A & 2009 & ${\bf 3.31} \pm {\bf 0.03}$ \\
    MACRO    & B & 2002 & ${\bf 3.22} \pm {\bf 0.08}$ \\
    Borexino & C & 2012 & ${\bf 3.41} \pm {\bf 0.01}$ \\
   \end{tabular}
\end{center}
\end{table}
 indicate variations due to systematics.  For the present simulation,
 the value given by Borexino was adopted, as this detector is located
 in Hall C next to the \darkside experiment.

\subsubsection{Mean energy and differential energy spectrum}

 The underground muon kinetic energy spectrum can be approximated
 by the following relation  \cite{macro02}:

 $$
    {dN\over{dE_{\mu}}} = const \cdot (E_{\mu} + \epsilon(1-e^{-\beta
    h}))^{-\alpha} 
   \label{eq:one}
 $$

 \hskip -\parindent
 In the above,  $E_{\mu}$ is the muon kinetic energy at slant depth
 $h$, and $\alpha$ is the surface muon spectral index.   For this
 approximation the quantities $\beta$ and $\epsilon$ are related to 
 muon energy loss mechanisms in rock, and are assumed constant.   From
 the above equation, the average muon kinetic energy at slant
 depth $h$ is:

 $$
    <\!E_{\mu}\!> \, = \, {\epsilon(1-e^{-\beta h})\over{\alpha -2}}
 $$

 \hskip -\parindent
 Experimental results for the spectral index $\alpha$ and the mean
 muon energy for LNGS were reported by MACRO \cite{macro02} and are
 reproduced in Table~\ref{index}:

   \begin{table}[hbt]
    \begin{center}
     \caption{\label{index} The mean energy of single and double muon
  events as measured by MACRO}
   \begin{tabular}{lcl}
     event type  & mean muon energy (GeV)                        &
     spectral index $\alpha$ \\ 
     \hline
     single muon & 270 $\pm$ \, 3 $_{(stat)}$ $\pm$ 18 $_{(syst)}$ &
     3.79 $\pm$ 0.02 $_{(stat)}$ $\pm$ 0.11 $_{(syst)}$ \\ 
     double muon & 381 $\pm$ 13 $_{(stat)}$ $\pm$ 21 $_{(syst)}$   &
     3.25 $\pm$ 0.06 $_{(stat)}$ $\pm$ 0.07 $_{(syst)}$ \\ 
     \end{tabular}
     \end{center}
   \end{table}

 \hskip -\parindent
 The functional description of the energy spectrum given by equation
 (\ref{eq:one}) permits direct sampling since it can be integrated
 and inverted in analytic form.  Adopting the values 
 $\epsilon = 0.392 \times 10^{-3}$ and $\beta = 635$~GeV this
 procedure reproduces the measured mean energy for single and double
 muon events, and gives an overall mean energy of 283\,$\pm$\,19 GeV
 for cosmogenic muons at LNGS.

\subsubsection{Bundles}

\begin{figure}[bht]
    \centering
    \includegraphics[width=3.2in]{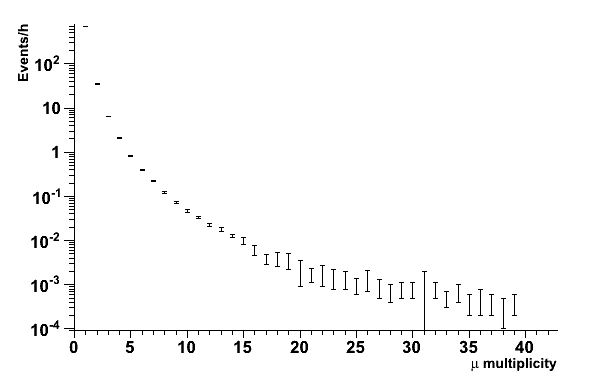}
    \includegraphics[width=3.2in]{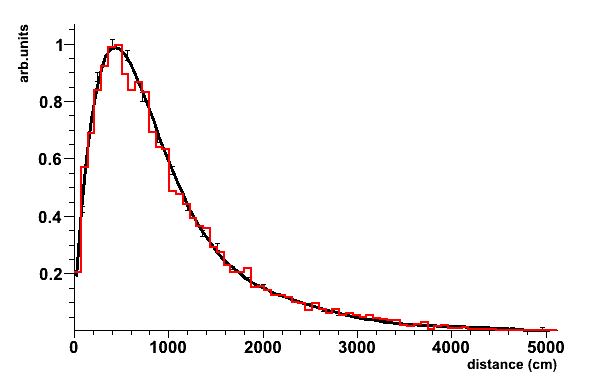}
    \caption{ (a) Multiplicity of muons recorded by the MACRO
              detector for cosmogenic muon events. 
               \,\, (b) Spatial separation of muons obtained for
              cosmogenic events in the 
              MACRO detector featuring two coincident muons (red data
              points). The black line shows 
              a simple higher order polynomial fit used to implement
              sampling of the distribution. }
   \label{fig:bundle}
   \end{figure}

 Muon bundles were investigated at LNGS by the LVD and MACRO experiments.
 The experimental results used here were taken from MACRO~\cite{bundles}.
 Figure \ref{fig:bundle}  shows the measured muon multiplicity and the
 spatial separation between muons for double muon events.   A
 simulation to study muon multiplicity used simplified
 sampling of the distribution up to a muon multiplicity of 4.   The
 distance between muons within a bundle was chosen according to the
 distribution measured for double muon events, and all muons within a
 bundle are given the same direction. 

 The effect of multi-muon events for the Borexino detector geometry
 was evaluated assuming the measured multi-muon event rate of
 about 6\% from MACRO.   Approximately 1.5\% of muon events in Borexino feature
 more than one muon.   In addition, about 12\% of the single muons
 crossing the Borexino inner detector belong to multi-muon events.

\subsubsection{Event generation}

 In order to implement the simulation, a combination of azimuthal 
 and zenith angles is selected according to the measured muon angular 
 distribution.   A map of the LNGS (Gran Sasso mountain) overburden,
 which was prepared 
 by the MACRO collaboration \cite{PhysRevD.52.3793,max}, is used to
 translate the muon direction into the respective slant depth $h$. 
 Next, the muon event type is set to either a single muon or muon bundle event.
 In the case of muon bundles, a multiplicity of up to 4 muons is
 sampled from the measured multiplicity spectrum.   The probability
 for muon events with larger multiplicity is less than 0.2\% and these
 events are treated as  muon bundles of multiplicity 4 to simplify the
 calculation.

 Finally, the muon kinetic energy is selected as a function of slant
 depth $h$ and muon event type by sampling from the parameterized
 single or double muon event spectrum.   The latter is used for 
 muon bundle events of all multiplicities since no experimental
 information is available for events with higher muon multiplicities.
 About 1.4\% of the cosmogenic events at LNGS contain more than
 2 muons.
 
 The muon events are randomly placed on a plane located above the
 hall geometry.  A sufficiently large
 plane above the hall is considered to fully illuminate the hall
 including the detector.  Other muon tracks outside this envelope
 are rejected.

 \subsubsection{Angular distribution}

 The slant depth of the muon transit through the rock overburden depends
 on both azimuthal and zenith angles. The azimuthal angular dependence
 of the slant depth in rock, $h \,\, (g \, cm^{-2})$, depends on the profile
 of the Gran Sasso mountain covering the experimental cavern.   Hence,
 both the intensity and energy profiles of the muons are a function of
 their incident direction on the detector.

 \begin{figure}[thb]
   \centering
   \includegraphics[width=3.2in]{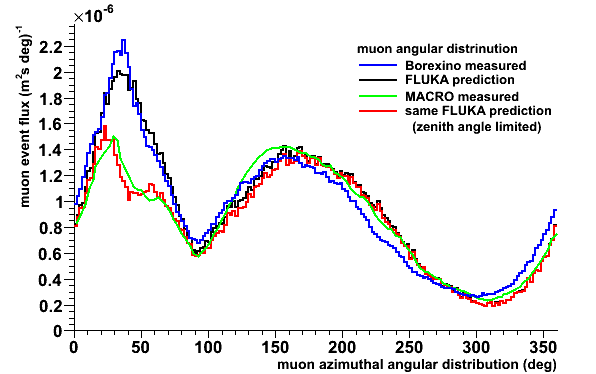}
   \includegraphics[width=3.2in]{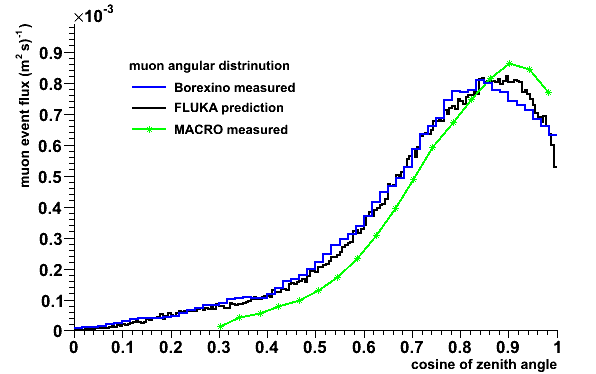}
   \caption{Muon azimuthal (a) and zenith (b) angular
             distribution at LNGS for polar coordinate system pointing
             up and North with clockwise increasing angle.
             Blue line: Borexino data,  green line: MACRO data,
             black (red) line: \Fluka predictions (zenith angle limited).
            }
   \label{fig:angular}
   \end{figure}

 A measurement of the cosmogenic muon angular distribution at LNGS was
 performed by MACRO~\cite{macro93} for Hall B.  Azimuthal and zenith
 projections of the distribution are shown by the green histograms in
 Figure~\ref{fig:angular}. 
 The distributions are compared to recent results from
 Borexino~\cite{cosmogenic_bx} for Hall C given by the blue histograms.
 The difference in the experimental azimuthal spectra for angles near
 45 degrees is caused by the MACRO zenith angular acceptance limit of
 about 60 degree.   The data are compared to a \Fluka simulation which
 traced muons, initiated by cosmic rays in the upper atmosphere, to the
 experimental halls~\cite{max}.   The spectra were normalized to the
 Borexino measured total muon flux.   The predictions for full detector
 acceptance are shown by the black histograms in Figure~\ref{fig:angular}.
 Imposing the zenith angle limitation of MACRO in the simulation
 reproduces the same feature as found in the data.   The resulting
 azimuthal spectrum is shown by the red histogram.   Good agreement is
 found between data and predictions.  The small visible shift in the
 azimuthal distribution is due to the change of location for the two
 experiments.

 \subsection{Muon-induced secondaries}

 \subsubsection{Geometry}

 The cosmogenic radiation field at deep underground sites is composed
 of muons and muon-induced secondaries.   Incident muons are allowed
 to develop particle showers as they pass through a 700~cm thick
 layer of Gran Sasso rock \cite{PhysRevD.52.3793} surrounding all
 sides of Hall C.   The rock thickness to fully develop the shower
 was determined by simulation, see Figure~\ref{fig:nEkin}a.
 However for computational
 purposes, this thickness was divided into three sections to permit
 simulation of electromagnetic processes with increasing detail as
 the shower approached the cavern.  
 The basic assumption of this
 procedure is to implement an initially homogeneous distribution
 of muons at every space point within the rock of appropriate
 direction and energy.

 \subsubsection{Propagation through rock}

 The cosmogenic radiation which is present at the cavern walls is
 approximated in two separate steps to reduce computation time.  
 In the first step, the muon radiation field as described in the
 previous section, is considered.   These muons were located on the
 outside of the rock layer surrounding Hall C, and allowed to propagate
 without interactions through the rock and recorded if they
 entered the hall. At these high energies, the muon trajectory is
 approximately unchanged by interactions. Muons entering the hall were
 then propagated 
 a second time with all physics processes enabled to derive a
 description of the cosmogenic radiation field including all
 muon-induced secondaries.   All secondaries are treated as part
 of the event.

 \subsubsection{Connection of simulation to measured muon event flux}

 The muon flux, $\Phi_{sim}$, inside the cavern expressed per simulated
 muon event, is determined by the simulation as described
 above.  An empty spherical volume was inserted inside the cavern, and exposed
 to the muon radiation field in order to determine the flux incident
 on the Borexino detector. The muon fluence estimated by the track
 length-density inside the sphere was obtain by a standard \Fluka
 scoring option.   Choosing a spherical volume to obtain the muon
 fluence correctly accounts for the angular distribution of
 the muons.      Moreover, the fluence through a sphere is a direct 
 estimator for the flux through the cross sectional area of the sphere.

 The length of the time-period considered in the simulation (lifetime),
 is given by the number of simulated muon events compared to
 the ratio of the simulated to measured total muon flux,
 $\Phi_{exp}$, from Table~\ref{mu_flux}.

 $$ T [s] = N_{events} \cdot { \Phi_{sim} [events^{-1} cm^{-2}] \over{
 \Phi_{exp} [s^{-1} cm^{-2}] }} $$


 \section{Validation}

The \Fluka predictions using Borexino as a model, 
are now compared to available data from
 similar experiments at comparable depth with a focus on the more recent
 measurements.   These include a new analysis from the LVD
 experiment~\cite{selvi1,selvi2} located in Hall A at LNGS and 
 results available from the KamLAND detector~\cite{kamland} which is
 operated at the Kamioka mine at a similar depth.

 Available information about muon-induced secondaries deep underground
 is limited to measurements of cosmogenic neutrons and isotopes in
 large liquid scintillator detectors.  As an example, a realistic 
  model of the Borexino
 experiment~\cite{bx_muons,bx_paper} was implemented in \Fluka in order
 to validate a FLUKA simulation of neutron production. The Borexino
 detector features a well shielded, 
 large volume liquid scintillator and provides high quality measurements
 of cosmogenic muon-induced neutrons with reduced systematic
 uncertainties in addition to the available results for the
 cosmogenic muon flux and the $^{11}$C production rate.

\subsection{Detector geometries}

  \begin{figure}[bht]
   \centering
   \includegraphics[height=3.0in]{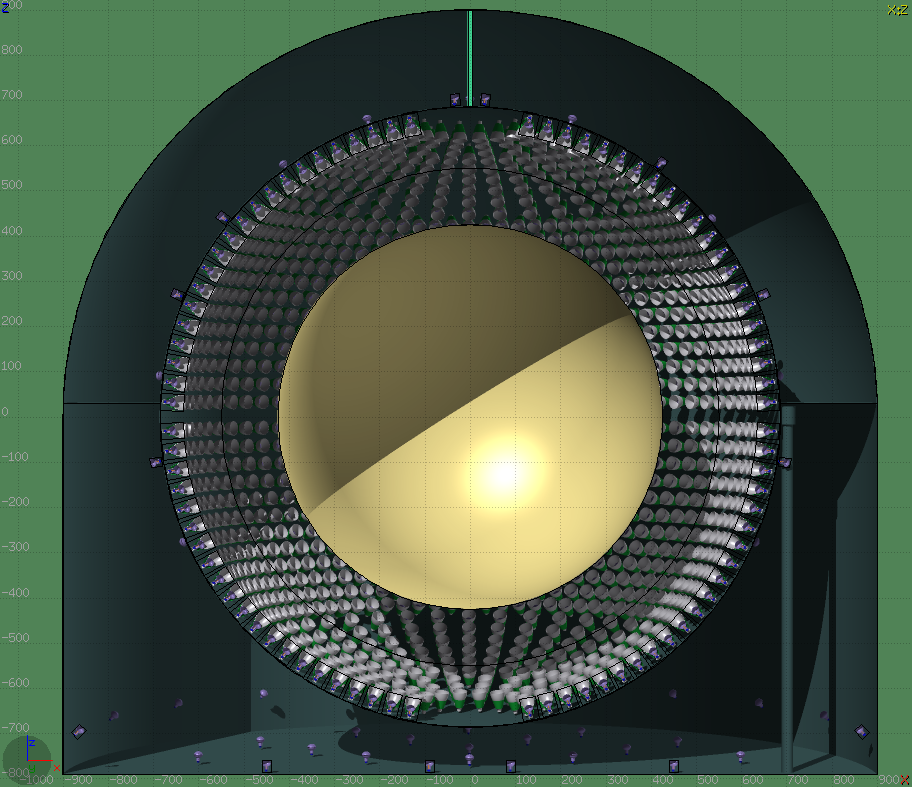}
   \caption{ A view of the \Fluka implemented Borexino detector geometry.
             The sensitive volume of the Borexino detector, the Inner
             Vessel, is shown by the golden sphere.  It is centered inside
             two (transparent) buffer regions which are contained inside
             a stainless steel sphere which supports the optical modules.
             This inner detector is placed inside a domed, cylindrical
             water tank which functions as a \v{C}erenkov detector to identify
             and help track cosmogenic muons.
            }
  \label{fig:bx}
  \end{figure}

A cross-sectional view of the geometry implemented for the Borexino
 experiment is given in Figure~\ref{fig:bx}.  The central sensitive
 liquid scintillator region of radius 425~cm having a mass of
 approximately 278 tons is contained in a spherical, nylon inner
 vessel.  It is located at the center of a 685~cm radius
 Stainless Steel Sphere (SSS) filled with inert liquid scintillator
 as a passive shield.  The SSS in turn is placed inside a 900~cm
 radius domed, cylindrical water tank providing additional shielding
 which functions as a water \v{C}erenkov muon veto detector.  The material
 of the detector components are carefully modeled in the simulation 
according to available information for Borexino.  The liquid
 scintillator is Pseudocumene with the addition of a 1.5 g/l
 fraction of wavelength shifter 2,5-diphenyl-oxazole (C$_{15}$H$_{11}$NO). 
The \Fluka simulation also implemented the photo-sensors of the
 inner detector, but no attempt was made to include optical transport and
 detector effects or to address the creation of realistic detector signals.
 Instead, the production of cosmogenic isotopes as well as neutron capture
 reactions on hydrogen inside the Inner Vessel were recorded.

Available data on muon-induced neutrons in underground
 laboratories before 2007 was summarized in Ref.~\cite{russian}.   It should
 be pointed out that this report includes most of the available data
 and that these data were obtained with similar systematic uncertainties.
 In addition to the data, more recent results from the LVD experiment
 are available. These new results are based on an analysis making use
 of a detailed simulation of the full LVD detector~\cite{selvi1}.
 The new results differ significantly from earlier published values for
 the same experiment.   Further, experimental data is now available from
 the KamLAND experiment.   The KamLAND detector is very similar in design
 to Borexino and situated at a comparable depth.  It has a reported average
 muon energy of $260 \pm 8$~GeV \cite{kamland}.

 Both measured and
 \Fluka simulated results are presented for the neutron yield and
 cosmogenic isotope production.   A measurement of the $^{11}$C
 production in liquid scintillator was also recently published by the Borexino
 experiment ~\cite{bx11c}.   All experimental information is
 based on neutron capture measurements in liquid scintillator detectors.


 \subsubsection{Total muon-induced neutron yield}

\begin{figure}[bht]
   \centering
   \includegraphics[height=2.9in]{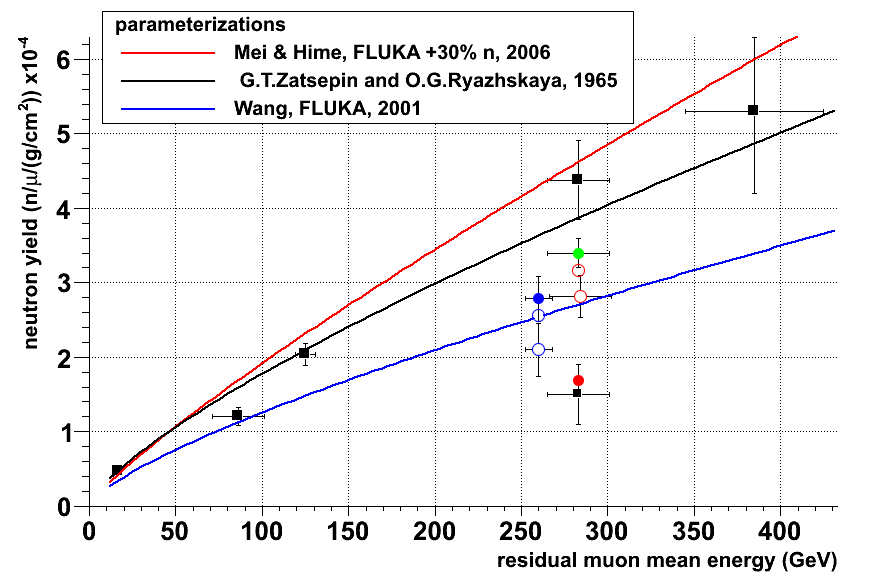}
   \caption{ Comparison between experimental data and simulation for
             muon-induced neutron yield in liquid scintillator as a
             function of mean muon energy:  from report~\protect{\cite{russian}}
             (black), most recent LVD (solid green), KamLAND (solid blue)
             and CTF (solid red).   The open circles
             are \Fluka predictions, see text for details.  Different
             suggested parameterizations of the neutron yield are shown
             by the thin lines: from report~\protect{\cite{russian}} (black),
             Mei and Hime~\protect{\cite{hime}} (red) and Wang~\protect{\cite{wang}} (blue).
           }
  \label{fig:yield}
  \end{figure}

 The most measured quantity for studying cosmogenic backgrounds is
 the total muon-induced neutron yield in liquid scintillator.
 Figure~\ref{fig:yield} gives a summary of the reported neutron yield
 as a function of mean muon energy.   The early measurements described
 in Ref.~\cite{russian} are shown by the solid black symbols together with
 a prediction (black line) which predates most measurements.   Two
 values are given for the LVD experiment which differ by more than a
 factor of 2.  The earlier lower yield was later explained by an
 additional energy threshold on the cosmogenic neutrons~\cite{lvd05}.

 The two LVD results were followed by a new analysis making use of a
 full detector simulation shown by the solid green
 symbol\footnote{The evolution of the LVD result for the cosmogenic
 neutron yield is indicative of the challenges to perform this
 measurement. The assigned systematic experimental uncertainties
 should be considered somewhat optimistic for older data points.}.
 Another measurement of the neutron yield at LNGS was reported
 by the Borexino collaboration (solid red symbol) just before the
 first LVD result was made public.   The detector volume was rather
 small and only single cosmogenic neutrons per crossing muon could be
 recorded due to the data acquisition system~\cite{galbiati}.

 The neutron yield as measured by the KamLAND experiment is shown by
 the solid blue symbol. It is compared to the reported \Fluka
 predictions given by the open blue symbol.  However, the
 simulated neutron yield can be corrected {\it a posteriori} for the
 incorrectly simulated rate of $^{11}C$ production as will be
 shown in the next subsection.  This correction increases the
 predicted neutron yield which is also indicated in 
 Figure~\ref{fig:yield}.
 Note that the \Fluka results for KamLAND were obtained with a simplified
 muon-induced radiation field and detector description.   Additionally,
 the reduced neutron count resulting from the missing deuteron
 interaction model in \Fluka and possible subsequent spallation were 
not addressed.
 
 The open red symbols are the preliminary results for the neutron yield
 obtained from the current \Fluka simulation.   The neutron yield after
 applying the same {\it a posteriori} correction is also indicated.
 The \Fluka predicted neutron yield in liquid scintillator material at
 energies close to the mean muon energy for LNGS of 283\,$\pm 19$\,GeV
 is in good agreement with more recent experimental data.
 Also shown in the Figure~\ref{fig:yield} are two previously suggested
 parameterizations derived by using \Flukadot.   The blue curve was given
 by Wang~\cite{wang} while the red curve is taken from the work of
 Mei and Hime~\cite{hime} which used an extra 30\% renormalization of
 the cosmogenic neutrons.

 \subsubsection{Cosmogenic isotope production}

 Radioactive isotopes produced by cosmic radiation are an
 important background to low-energy solar neutrino spectroscopy experiments
 like KamLAND and Borexino.   Of the produced radioactive
 isotopes, $^{11}$C poses the largest challenge because of its relatively
 long mean lifetime of approximately 30 minutes.   Therefore, the $^{11}$C
  \begin{figure}[tbh]
   \centering
   \includegraphics[height=2.9in]{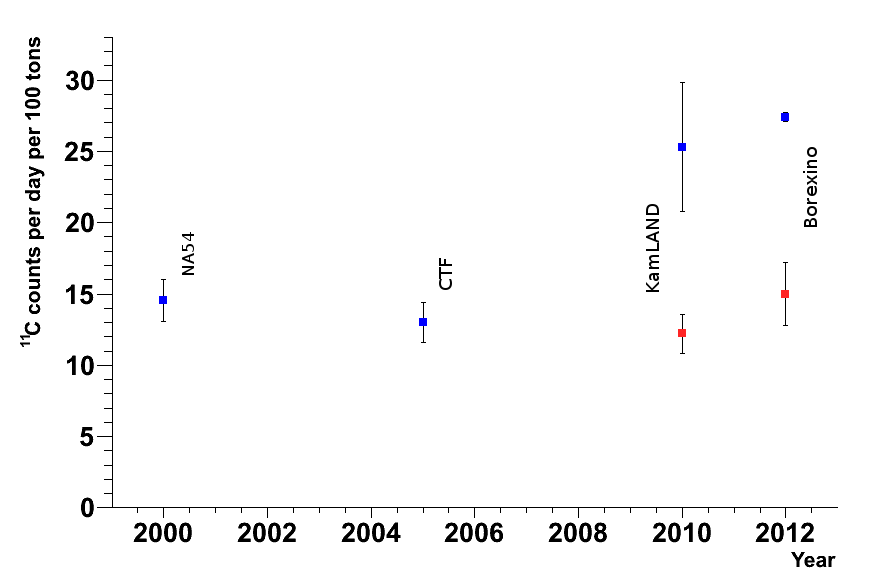}
   \caption{ Cosmogenic $^{11}$C production  per day for the nominal
             100 ton fiducial volume of the Borexino detector. Rates
             derived by scaling from earlier available experimental data
             (CERN experiment NA54~\protect{\cite{hagner}} and CTF~\protect{\cite{bxCTF}})
             are compared to the recently published
             $^{11}$C production rate (solid blue symbols) measured
             by KamLAND and Borexino.  Error bars show the reported
             experimental uncertainties.  Also shown are \Fluka predictions
             for the $^{11}$C production rate (solid red symbols) from
             KamLAND and the current simulation.
            }
  \label{fig:c11}
  \end{figure}
 production rate was studied extensively.  The current situation is depicted
 in Figure~\ref{fig:c11} as a function of time.

 As can be seen, the \Fluka predicted $^{11}$C production rate agrees with
 early measured results.   However, newer and more reliable measurements
 indicate a production rate about twice as large.   Because of missing
 details when simulating the Fermi-breakup of $^{12}$C in the current
 version of \Flukadot,  see subsection 2.4.2,  a significant underproduction
 of $^{11}$C is now expected.

 The measured production yield reported by KamLAND for $^{11}$C of
 $ 0.866 \pm 0.153 \times 10^{-4} (\mu \, g/cm^2)^{-1}$ is about 1/3 of
 the neutron production yield of $ 2.787 \pm 0.311 \times 10^{-4}
 (\mu \, g/cm^2)^{-1}$.  Further, KamLAND confirmed that about 96\% of
 the production of $^{11}$C proceeds through the photoproduction
 reaction  $^{12}$C($\gamma$,n)$^{11}$C  which produces a neutron.
 Consequently, the under-prediction of $^{11}$C
 production by \Flukadot, results in a reduced neutron yield.   Simple
 {\it a posteriori} scaling permits a cursory correction of the \Fluka
 predictions for the neutron yield based on the measured and simulated
 $^{11}$C production rates for the KamLAND and Borexino simulations.

 The under-predicted rate of $^{11}$C production for liquid scintillator
 material is responsible for a large fraction of the reported problems
    with the number of cosmogenic neutrons predicted by \Flukadot.   It is
 important to recognize the source of this missing component.  The class
 of missing neutrons from the simulation are not high energy neutrons
 as they are produced by photo-nuclear interactions from the significant
 flux of real photons in the muon radiation field.


 \subsubsection{Distance between neutron capture and muon track}

\begin{figure}[htb]
   \centering
   \hskip -2.5mm

   \includegraphics[width=3.2in]{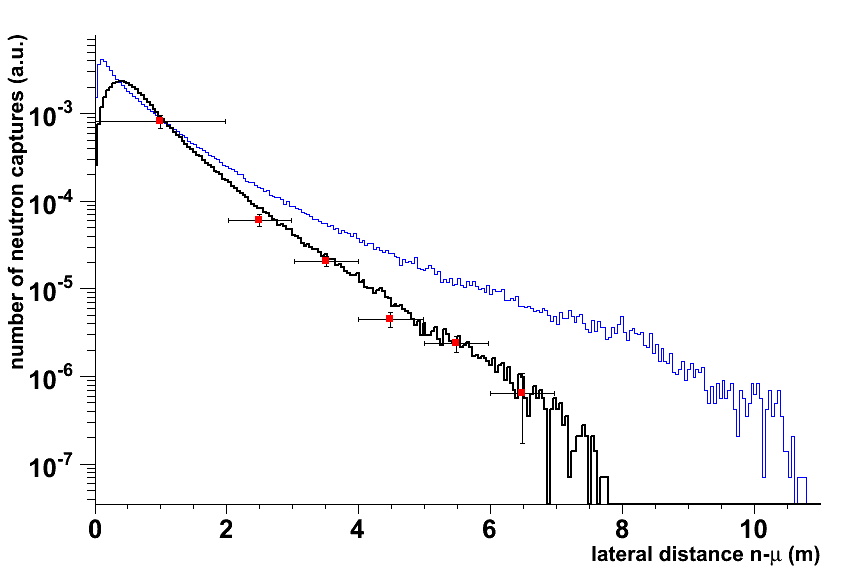}
   \includegraphics[width=3.2in]{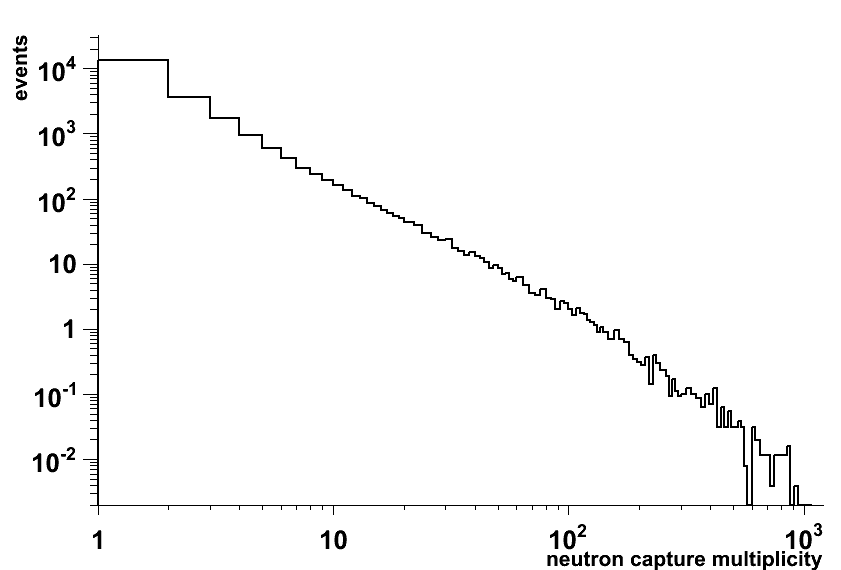}

   \caption{ (a) Comparison between experimental data (solid blue symbols)
             from LVD~\protect{\cite{lvdlat}} and \Fluka predictions
             (black histogram) for the distance between neutron capture
             and the muon track. \, \, (b) Muon-induced cosmogenic neutron
             multiplicity as predicted by \Flukadot.   Events with neutron
             multiplicities of several hundreds of neutrons per event
             are expected.
           }
  \label{fig:lateral}
  \end{figure}

 The lateral distance distribution of the neutron capture location
 from the parent muon track is shown in Figure~\ref{fig:lateral}a.
 The thin blue histogram reflects the \Fluka prediction of the
 distribution for the Borexino detector.  The shape and extent of
 the distribution is biased by the geometry and size of the
 detector.   The thick black histogram is obtained by folding the 
 simulated distribution with the muon track reconstruction
 uncertainties~\cite{bx_muons}.   In addition, both the distance of
 the muon track and the neutron capture point from the detector
 center were restricted to less than 400 cm.  The effect of the muon
 track uncertainties is most visible for neutron captures close to
 the parent muon track while the restriction in distance from the
 detector center impacts the shape and extent of the distribution.

 The shape of the predicted distribution is compared to available data
 from the LVD experiment~\cite{lvdlat} (solid red symbols).   The
 experimental distribution is well reproduced by the simulation when
 the effective size of the Borexino detector is limited to a radius of
 400 cm.


 \subsubsection{Neutron multiplicity}

 Figure~\ref{fig:lateral}b presents the cosmogenic neutron multiplicity
 distribution as predicted by FLUKA for the Borexino detector.   Some events
 with very large multiplicities are found.   No experimental spectrum for
 the cosmogenic neutron multiplicity distribution is available.   A
 measurement of the neutron multiplicity is expected from the Borexino
 experiment in the near future.



\section{Cosmogenic background predictions for \darkside}

 The \darkside detector will be installed inside the Counting Test Facility
 (CTF) which is located in Hall C at LNGS adjacent to the Borexino detector.
 Because of the proximity of the two experiments, the same muon induced
 cosmogenic radiation field which was used for the comparison to available
 experimental data can also be used to simulate the cosmogenic
 background for \Darkside.

 \subsection{Detector geometry}

\begin{figure}[bht]
    \centering

    \includegraphics[width=2.6in]{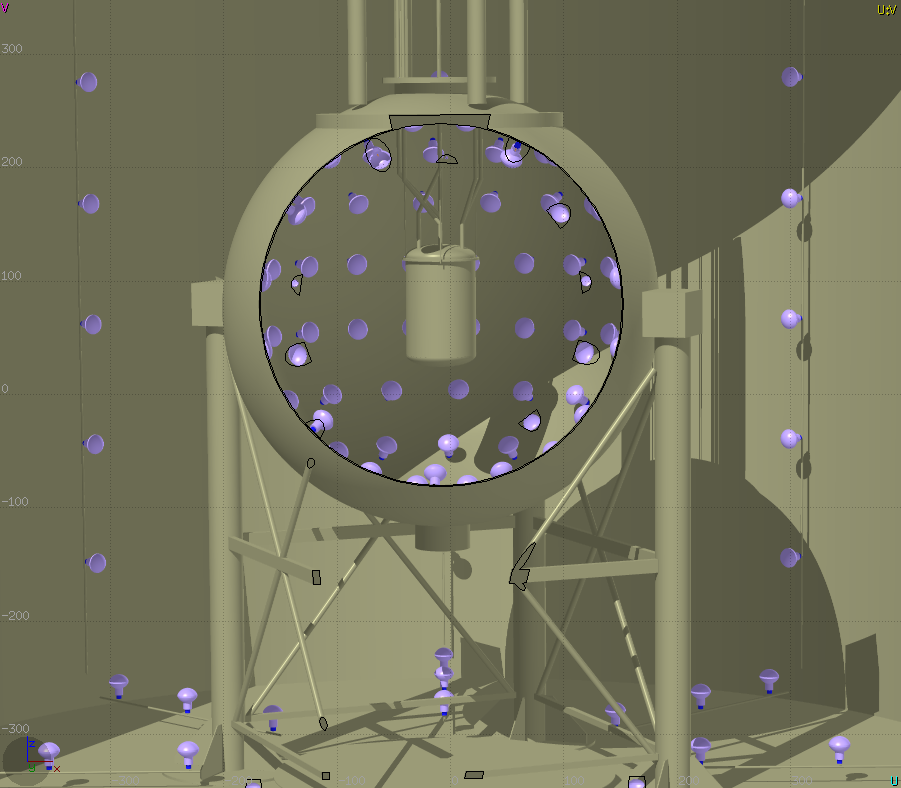}
    \includegraphics[width=2.6in]{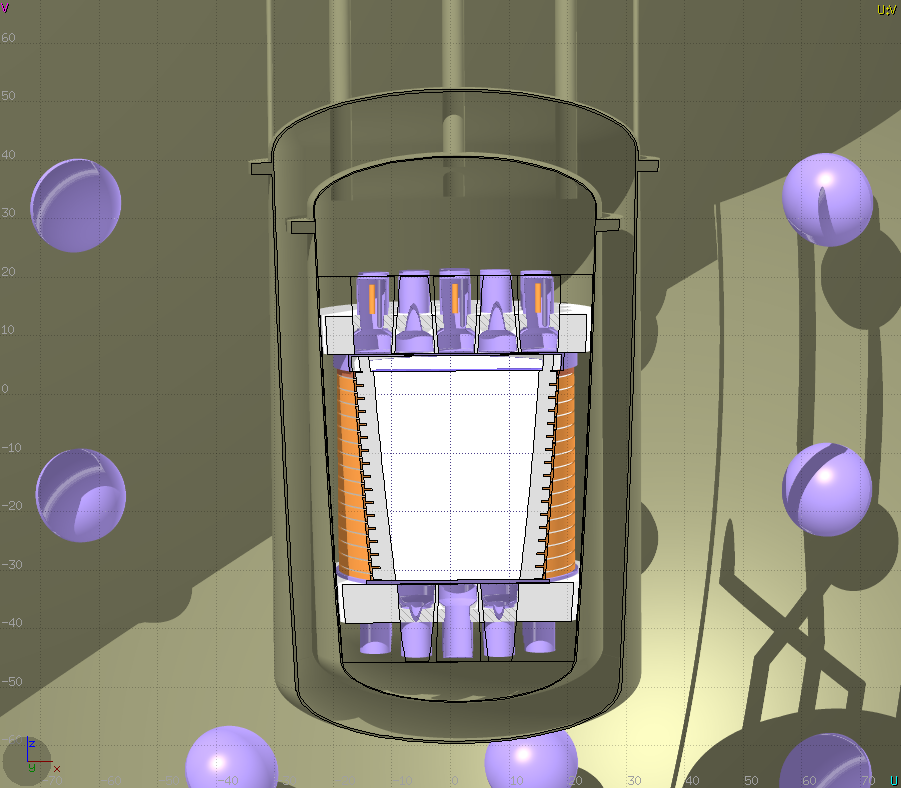}
    \vskip 1mm
    \includegraphics[width=2.6in]{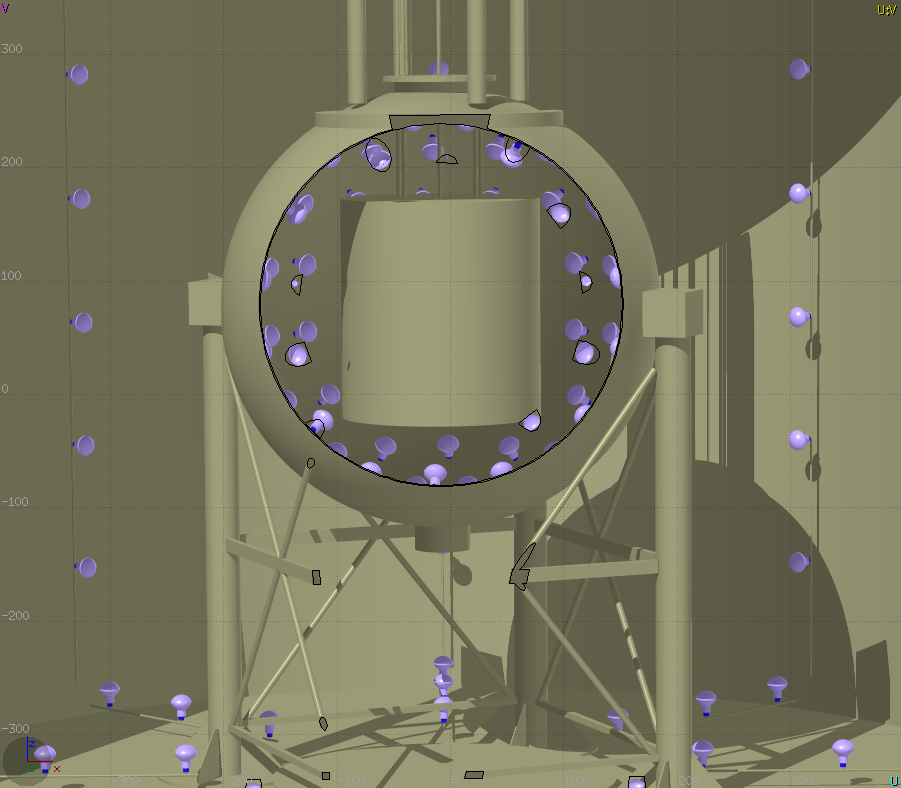}
    \includegraphics[width=2.6in]{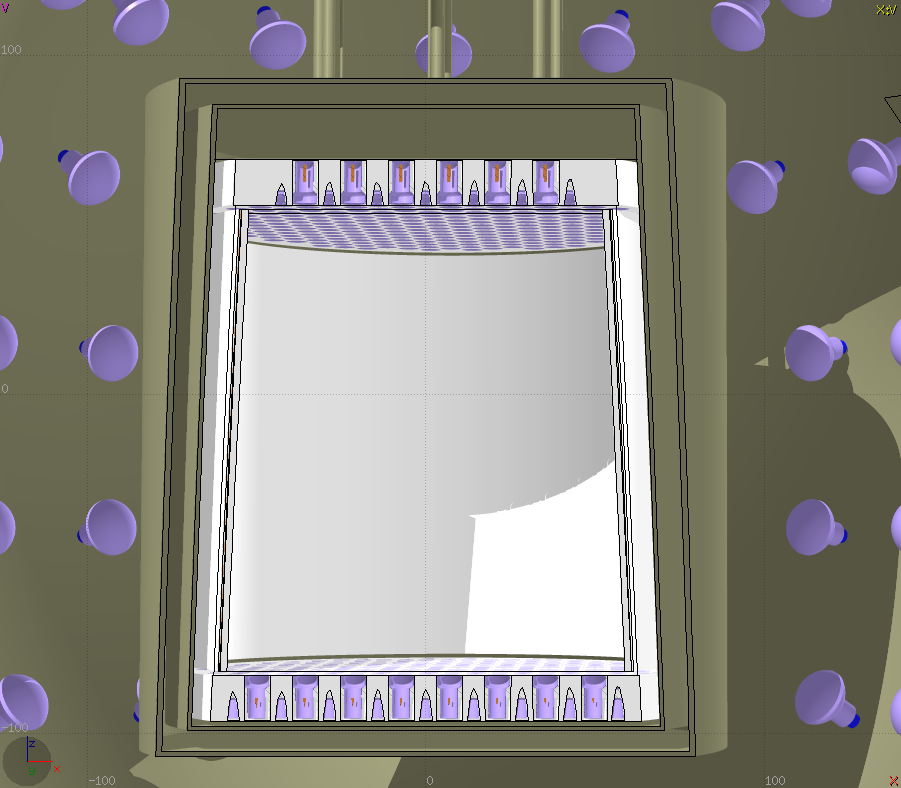}

    \caption{ \Fluka implementation of the \darkside experiment. 
              The Dewar containing the TPC is shown mounted inside
              the LSV on the left while the graphs on the right
              expose details of the geometry implemented for the
              TPC.   Top row: \Darkside-50; \,\,
                     Bottom row: \Darkside-G2.
            }
   \label{fig:outer}
   \end{figure}

 The \darkside detector consists of a cylindrical, two-phase underground
 liquid argon Time Projection Chamber (TPC)~\cite{ds50}.
The TPC is contained inside
 a thin-walled stainless steel Dewar.   The sensitive region is viewed by
 photomultiplier tubes from the top and bottom.   It is surrounded by a
 TPB coated Teflon cylinder which acts as an optical reflector and wavelength
 shifter.
 The Teflon cylinder also supports a set of copper rings which provide the
 electric field.  The \Darkside-50 experiment with an active mass of 50 kg
 is currently under construction.  It features a sensitive volume of
 about 35 cm diameter and 35 cm height.   The inner detector will be
 immersed in a highly efficient, borated Liquid-Scintillator neutron
 Veto (LSV) to reduce external backgrounds and to monitor cosmogenic and
 radiogenic neutron backgrounds {\it in situ}.  To further reduce cosmogenic
 backgrounds, the LSV is surrounded by ultra pure water inside
 a large cylindrical water tank (CTF) which functions
 as muon veto \v{C}erenkov detector.  The CTF will 
 also provide a passive shield against low energy external backgrounds.
 The implementation of \darkside in FLUKA is shown in
 Figure~\ref{fig:outer}.

 The \darkside experiment is in addition to its scientific reach, also 
a prototype for the development of a
 ton-sized, next generation TPC, as the
 design of the veto detector system permits the upgrade to \Darkside-G2
 with a sensitive mass of approximately 3.3 tons of underground liquid argon.
 The ton-sized TPC features a sensitive volume of about 150 cm diameter
 and 135 cm height.  In general, the rate of cosmogenic neutron background
 scales with the size of the sensitive detector volume and is
 substantially larger for \Darkside-G2 compared to that of \Darkside-50.
 Similarly, the increased size of the TPC displaces a larger amount
 of liquid scintillator inside the LSV and reduces the efficiency of
 detecting neutrons.


 \subsubsection{Counting Test Facility}

 The CTF~\cite{ctf} is modeled in this simulation 
as a cylindrical tank of 11 meter diameter
 and 10 meter height constructed from 8 mm thick carbon steel.   It is
 filled with ultra pure water and placed on an additional 10 cm thick
 layer of steel.   The inside tank surfaces are
 covered by sheets of a special type of layered Tyvek foil~\cite{tyvek}
 to enhance light collection.   For the relevant spectral range determined
 by the produced \v{C}erenkov light and the sensitivity of the PMTs, a
 reflectivity in water of greater than 95\% is expected.   In the simulation a
 more conservative constant reflectivity of 80\% was assumed.   \v{C}erenkov
 light produced in the water is measured by 80 ETL-9351 eight-inch PMTs which
 are placed along the floor and the walls of the CTF.   The production,
 propagation and detection of the \v{C}erenkov photons in the CTF is simulated
 in detail by FLUKA taking into account a constant refractive index of
 1.334 and a realistic absorption spectrum for pure water \cite{pureH2O}.
 The photo sensors are modeled with their measured quantum efficiency
 spectrum scaled by a 70\% photoelectron      collection efficiency
 \cite{McCarty}.  A study to evaluate the placement of the PMTs indicated
 only a weak dependence of the overall efficiency to detect muons as a
 function of the spatial distribution of the optical sensors.   
The insensitivity to the spatial dependence is attributed
 to the low absorption in pure water combined with the high surface
 reflectivity.

 \subsubsection{Liquid-Scintillator neutron Veto}

 The LSV \cite{lsv} is implemented as a 4 meter diameter stainless steel
 sphere with 8 mm wall thickness.  It is filled with a borated liquid
 scintillator consisting of a 1:1 mixture of Pseudocumene and
 Trimethylborate.   The sphere is located inside the CTF and both the
 inside and the outside surfaces are covered by the Tyvek reflector.
 The LSV is equipped with 110 low-background glass-bulb Hamamatsu
 R5912-HQE-LRI eight-inch  PMTs which are mounted on the sphere facing
 inward.  Optical processes inside the LSV were not simulated.   
Instead, the un-quenched raw energy
 deposition per event is recorded.

 \subsubsection{Inner detector}

 Both, the DarkSide-50 and the DarkSide-G2~\cite{ds50,dsg2}
 configurations of the inner
 detector were implemented in the FLUKA simulation.   Initial studies
 were carried out for the smaller DarkSide-50 design, however the
 following results are for the ton-sized DarkSide-G2 geometry.   
In this case, the cosmogenic
 background requirements are more stringent as DarkSide-G2 has a much
 larger sensitive volume.  Thus the results can be viewed as a
 conservative upper limit to those expected for DarkSide-50.   The
 geometry and material 
 composition of the Dewar and TPC for the  DarkSide-50 setup were
 modeled in detail according to the latest detector design drawings.
 This was then used to prepare a realistic, scaled model
 for the ton-sized configuration.   In the simulation the sensitive
 volume is viewed by  283 three-inch Hamamatsu,
 low-background R11065 PMTs positioned on the top and an equal number
 on the bottom of the TPC.   As for the LSV,
 no simulation of the optical processes inside the TPC were
 undertaken.
The number of particles entering the sensitive
 volume, their type and the total raw energy deposited per event were
 recorded.

 \subsection{Simulation results}

 Starting from full cosmogenic events which were ``frozen'' on the
 walls inside of the experimental Hall C at LNGS, all particles were
 transported by FLUKA through the section of the cavern which
 contains the \Darkside-G2
 experiment.   Results are presented based on a total number of simulated
 cosmogenic events corresponding to a lifetime of approximately 36 years.
 The statistical uncertainty of the results is of the order
 of a few percent, and in any event, is smaller than the systematic
 uncertainties.   Events for which at least one
 particle reached the CTF water tank were recorded as a first step.  
The predicted rate for
 these cosmogenic events at the outside of the CTF is approximately  3.45
 events per minute.   For about 23\% of the events, the original
 cosmogenic muon does not reach the CTF water tank.

 The complete \Darkside-G2 detector setup was then exposed to all events
 which were recorded at the outside of the CTF in a second step of the
 simulation.   All physics processes were turned on making use of the
 FLUKA defaults setting PRECISIO(n).   In this step of the simulation
 \u{C}erenkov photons were created inside the CTF water tank.   However,
 because of CPU considerations they were not initially transported.
 The rate of cosmogenic events with at least one particle reaching the LSV
 is predicted to be approximately  0.30  events per minute.   This
 reduction in rate is the result of both the smaller size of the
 volume and the passive shielding of the water tank.
 Similarly, the rate of cosmogenic events with at least one particle
 reaching the sensitive region inside the TPC is predicted by FLUKA to
 be  0.07  events per minute.   The decrease in rate is caused by the
 smaller size of the sensitive liquid argon region and the passive
 shielding of the liquid-scintillator.   Reported rates are upper
 limits since no energy deposition in the respective detectors was
 required.  The expected cosmogenic event rates are summarized in
 Table~\ref{rates}.

\begin{table}[thb]
 \begin{center}
\caption{\label{rates} Expected cosmogenic event rates}
   \begin{tabular}{lc}
    just outside of  & cosmogenic event rate (per minute) \\
    \hline
    CTF      &  3.45 \\     
    LSV      &  0.30 \\
    sensitive liquid argon region & 0.07 \\
   \end{tabular}
\end{center}
\end{table}

  \vskip 3mm

 In order to further study the predicted cosmogenic background to the
 \Darkside-G2 experiment the subset of events with at least one particle
 reaching the inner sensitive region was considered.   Events with
 more than 50 coincident particles reaching the sensitive volume were
 also rejected since these give a clear signal in the TPC and can be
 rejected.   When these cuts are applied, the fraction of cosmogenic
 events in which the original muon does not enter the CTF is reduced to 
 7.7  events per year.

\begin{figure}[htb]
    \centering
    \includegraphics[width=3.4in]{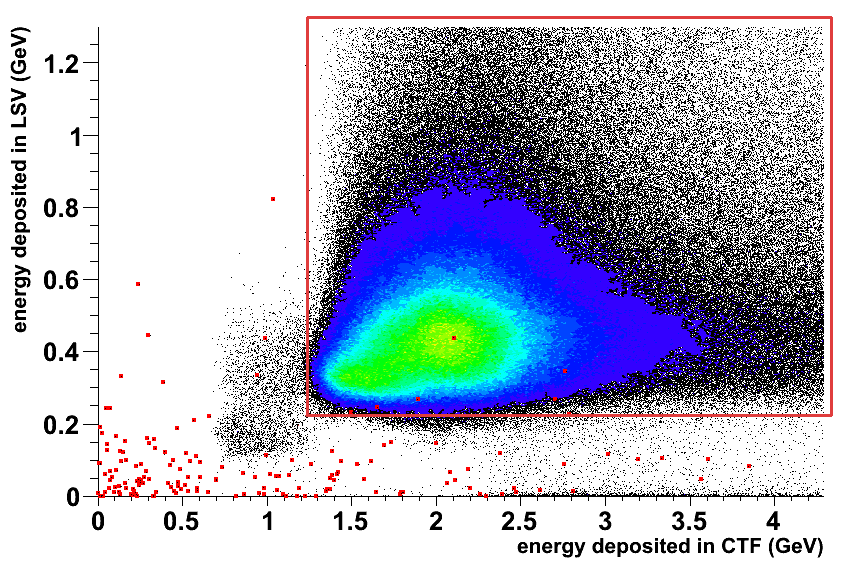}
    \caption{ Energy deposition inside CTF and LSV for cosmogenic events with
              at least one particle reaching the sensitive volume of the 
              inner detector.
              }
   \label{fig:dEdE}
   \end{figure}

 In Figure~\ref{fig:dEdE} the raw energy deposited inside the CTF and the LSV
 are graphed with respect to each other.   Cosmogenic events with at least
 one particle reaching the sensitive liquid argon volume almost always
 deposits  a significant amount of energy inside the veto detectors.
 The dominant region in the scatter plot is indicated by the red box and
 limited by  dE$_{(CTF)}$\,$>$\,1.3 GeV  and  dE$_{(LSV)}$\,$>$\,0.2 GeV. 
 This region
 corresponds to events with the original cosmogenic muon traversing both
 detectors.   The superimposed color contour plot indicates the shape of the
 most frequent energy deposition for cosmogenic events resulting from the
 ionization of the relativistic muons.   Events with similar energy deposited
 inside the CTF but with dE$_{(LSV)}$\,$<$\,0.2 GeV  indicate that 
 the original muon traversed
 the CTF but missed the LSV.   A smaller set of events in the
 region of  0.7 GeV\,$<$\,dE$_{(CTF)}$\,$<$\,1.3 GeV results from low energy
 comogenic
 muons which traverse the water tank but stop inside the LSV.   The most
 difficult cosmogenic events to veto are
 found close to the origin of the graph.   For these events, little energy is
 deposited both inside the CTF and the LSV.   Cosmogenic events predicted by
 FLUKA without a direct muon into the CTF are superimposed on the graph with
 solid red symbols.   Practically all cosmogenic events 
 which have  small energy deposition in both the CTF and LSV fall
 into a class of events with no direct muon entering the CTF.

  \vskip 3mm

 The most important background events for the direct dark matter search
 consist of neutron-induced recoils in the sensitive volume from undetected
 neutrons.   In the next step of the simulation, all events for which at least
 one neutron (but $<$ 50 coincident particles) reached the sensitive liquid
 argon region were reprocessed with full treatment of optical processes
 inside the CTF.   A total of 19735 of these events, or 581 events per year,
 are predicted by FLUKA.   Only the raw energy deposited inside the LSV and
 the sensitive region of the TPC are available in the current simulation.
 Therefore, conservative criteria were defined to select events which are
 considered detectable by the LSV:  dE$_{(LSV)}$\,$>$\,1 MeV, and for events which
 fall into the energy range of a possible dark matter signal in the TPC:
  0\,$<$\,dE$_{(TPC)}$\,$<$\,1 MeV~\cite{dsg2}.

  \begin{figure}[htb]
    \centering
    \includegraphics[width=3.2in]{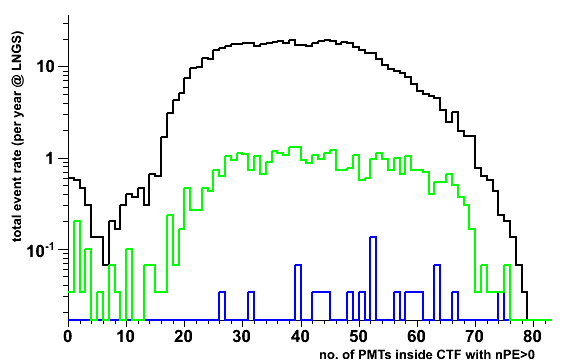}
    \includegraphics[width=3.2in]{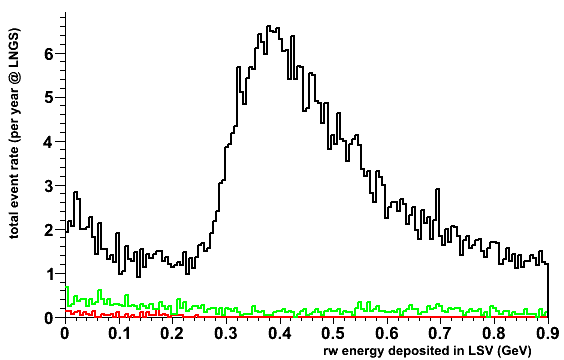}
    \caption{ (a) Number of PMTs with at least 1 registered photoelectron.
                  black: all events, green: raw dE inside TPC\,$>$\,0 and
                  $<$\,1 MeV, and blue: raw dE inside the LSV\,$<$\,1 Mev.
              \,\,
              (b) Un-quenched energy deposited inside the LSV. black:
                  all events, green: raw dE inside TPC $>$\,0 and $<$\,1 MeV,
                  and red: $<$\,10 PMTs with at least 1 registered
                  photoelectron.  
            }
   \label{fig:spectra}
   \end{figure}

 The predicted response of the veto detectors is shown in
 Figure~\ref{fig:spectra} 
 for an equivalent lifetime of approximately 36 years at LNGS.  The number of
 events found for the CTF as a function of PMTs which register a signal
 (one or more photoelectrons) is given by the black histogram on the left.
 The events shown by the blue histogram are found if the energy deposited
 inside the LSV is limited to less than 1 MeV.   The black histogram in the
 graph on the right           shows the predicted energy spectrum for the LSV.
 Limiting the sample to events with less than 10 PMTs registering a signal
 inside the CTF reduces the energy spectrum to the events shown by the red
 histogram.   The effect of selecting events in the energy range of interest
 for the TPC is shown for both distributions by the green histograms.

\begin{figure}[hbt]
    \centering
    \includegraphics[width=3.4in]{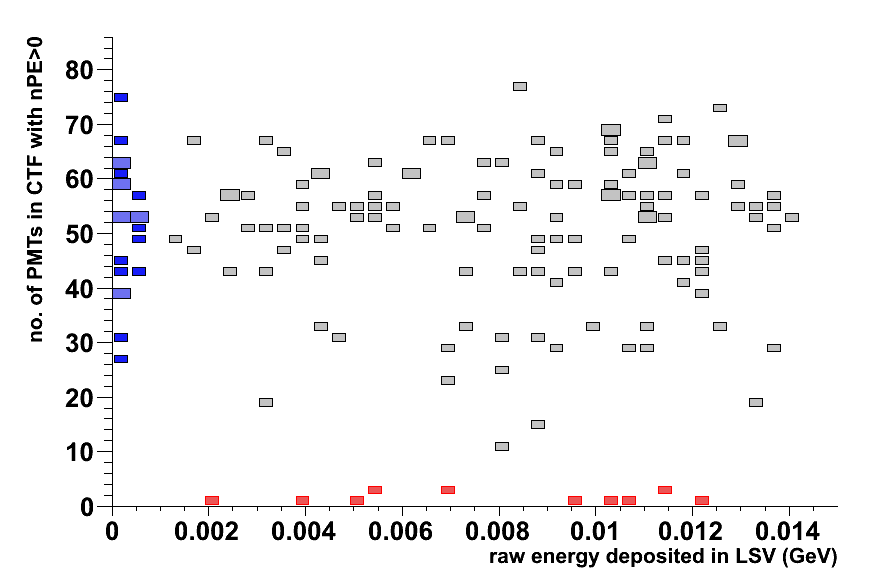}
    \caption{ Shown is the of number of PMTs which registered at least one
              photoelectron inside the CTF versus the un-quenched energy
              deposited in the LSV.  The red and blue color coded entries
              correspond to the respective events selected in
              Figure~\ref{fig:outer}.
            }
   \label{fig:2Dnew}
   \end{figure}

 The same information with focus on events with energy less than 14 MeV
 deposited in the LSV are shown in Figure~\ref{fig:2Dnew}.   The number of
 PMTs with a signal inside the CTF is graphed versus the raw energy
 deposited in the LSV.  Ten events with less than 3 PMTs recording a signal
 inside the CTF are considered to be missed by the muon veto and
 are colored red in the plot.   Similarly, twenty-one events with a raw energy
 deposition of less than 1 MeV inside the LSV are considered missed 
 and are colored blue in the graph.

 The FLUKA simulation predicts approximately 581 events per year in which at
 least one cosmogenic muon-induced neutron enters the sensitive liquid argon
 volume.   Out of these events, only 0.3 events per year fail to
 cause a signal in 3 or more PMTs of the CTF muon \u{C}erenkov veto.  At the
 same time, only 0.6 events per year deposit less than 1 MeV inside the
 LSV.   For a simulated live-time of approximately 36 years at LNGS, no event
 occurs where a neutron reached the sensitive liquid argon region but
 failed to trigger at least one of the two veto detectors.
  The rates reported are obtained without considering information
  from the TPC to further reject cosmogenic neutron events.


 The cosmogenic muon-induced neuton background rate to the \Darkside-50
 experiment is significantly smaller than for the much larger
 \Darkside-G2 configuration, while the same veto detectors are used
 for both implementations.   The rejection of external cosmogenic
 neutrons by the LSV increases for \Darkside-50 since there is more
 liquid-scintillator volume. In addition, this
 also increases the amount of passive shielding.


 \section{Conclusions}

 FLUKA predictions for cosmogenic muon-induced neutron backgrounds
 are found to be in reasonable agreement with available
 data. A few additional problems were identified and will be resolved
 in future versions of the code.  Corrections relating to the $(\gamma,
 n)$ cross section on $^{12}C$ for example, can be
 approximately included in the final results, as was discussed in the
 paper.  However, the discrepancy and suggested correction reported
 in Ref.~\cite{hime} were found invalid.

 The need of a detailed description of the full muon-induced
 radiation field for the considered underground site was described and
 shown to be important. Steps to explicitly prepare the cosmogenic
 radiation field with the FLUKA simulation package were described and
 the muon-induced radiation field which was
 used to carry out benchmark studies,     was then applied to the
 \darkside experiment. Background predictions for a direct dark
 matter search experiment were obtained for the next generation,
 ton-sized two-phase underground liquid argon detector.   It was
 found that the proposed dual active-veto system for the experiment
 provides sufficient shielding against cosmic radiation at the LGNS
 depth for a ton-sized \Darkside-G2 experiment for more than 5 years.

 Background levels for the \Darkside-50 experiment, which 
 use the same veto detector system, are expected to be significantly
 reduced because of the smaller size and the consequently increased
 volume of the liquid-scintillator shield and thus adequate for the
 smaller detector. In the future it is essential to monitor
 the cosmogenic neutron background levels, with focus on the veto
 detector system, in order to continue to benchmark the simulation. 
 Work is ongoing to extend the FLUKA studies to new, quality
 data on cosmogenic neutrons which will be available from the Borexino
 experiment.

 \section*{Acknowledgements}

 This work was supported in part by NSF awards 1004051 and 1242471.
 We would like to thank Alfredo Ferrari and
 Maximillian Sioli for FLUKA related help and acknowledge the use of
 FLAIR. We also would like to thank David D'Angelo, Quirin Maindl and
 Michael Wurm of the Borexino collaboration.


\bibliographystyle{unsrt}
\bibliography{references}

\begin{thebibliography}{10}

\bibitem{kolar}
V.~S. Narasimham.
\newblock {Perspectives of experimental neutrino physics in India}.
\newblock {\em Proc Indian Natn Sci Acad}, 70(A1):11--25, January 2004.

\bibitem{fluka1}
G.~Battistoni, S.~Muraro, P.R. Sala, F.~Cerutti, A.~Ferrari, S.~Roesler,
  A.~Fass\`o, and J.~Ranft.
\newblock {The FLUKA code: Description and benchmarking}.
\newblock {\em AIP Conference Proceeding}, 896:31--49, 2007.

\bibitem{fluka2}
A.~Fass\`o, A.~Ferrari, J.~Ranft, and P.R. Sala.
\newblock {FLUKA: a multi-particle transport code}.
\newblock 2005.

\bibitem{manual}
{FLUKA Collaboration}.
\newblock {FLUKA manual, ASCII or .pdf file available from FLUKA website and
  contained in FLUKA package}, November 2011.

\bibitem{fluka0}
{FLUKA Collaboration}.
\newblock {FLUKA official website, http://www.fluka.org}, 2012.

\bibitem{atlas}
A.~Antonelli, G.~Battistonii, A.~Ferrari, and P.~R. Sala.
\newblock {Study of radiative muon interactions at 300 GeV}.
\newblock {\em Frascati Physics Series, Calor 96}, VI:561--570, 1997.

\bibitem{fluka3}
A.~Fass\`o, A.~Ferrari, and P.~R. Sala.
\newblock {Designing Electron Accelerator Shielding with FLUKA}.
\newblock {\em Proc. 8th Int. Conf. on Radiation Shielding}, pages 643--649,
  April 1994.

\bibitem{ntof1}
C.~Borcea et~al.
\newblock {Results from the commissioning of the n\_TOF spallation neutron
  source at CERN}.
\newblock {\em Nucl. Instrum. Meth. A}, 513:524--537, 2003.

\bibitem{ntof}
nToF collaboration.
\newblock {TOF: Proposal for a neutron time of flight facility}.
\newblock {\em CERN-SPSC-99-8}, 1999.

\bibitem{sala}
the FLUKA~collaboration P.~Sala~et at.
\newblock {New Developments in FLUKA}.
\newblock {\em CERN Conference Proceeding, Varenna - 13th International
  Conference on Nuclear Reactions - FLUKA}, June 2012.

\bibitem{hime}
D.-M. Mei and A.~Hime.
\newblock Muon-induced background study for underground laboratories.
\newblock {\em Phys. Rev. D}, 73(5):053004, Mar 2006.

\bibitem{riznia}
R.~J. Jasim.
\newblock {\em {Monte Carlo Evaluation of Cosmogenic neutron background at Gran
  Sasso}}.
\newblock PhD thesis, University of Houston, 2011.

\bibitem{wulandari}
H.~Wulandari, J.~Jochum, W.~Rau, and F.~von Feilitzsch.
\newblock {Neutron background studies for the CRESST dark matter experiment}.
\newblock {\em arXiv HEP-EX/0401032}, 2004.

\bibitem{dementyev}
A.~Dementev, V.~Gurentsov, O.~Ryazhskaya, and N.~Sobolevsky.
\newblock {Production and transport of hadrons generated in nuclear cascades
  initiated by muons in the rock (exclusive approach)}.
\newblock {\em Nucl. Phys. Proc. Suppl.}, 70:486--488, 1999.

\bibitem{seasonal}
{Davide D'Angelo, for the Borexino collaboration}.
\newblock {Seasonal modulation in the Borexino cosmic muon signal}.
\newblock {\em Proceedings of the 32nd ICRC}, 2011.

\bibitem{decoherence}
{M. Ambrosio, the MACRO Collaboration}.
\newblock High statistics measurement of the underground muon pair separation
  at gran sasso.
\newblock {\em Phys. Rev. D}, 60:032001, Jun 1999.

\bibitem{macro02}
{M. Ambrosio, the MACRO Collaboration}.
\newblock Measurement of the residual energy of muons in the gran sasso
  underground laboratories.
\newblock {\em Astroparticle Physics}, 19:313--328, 2003.

\bibitem{bundles}
{J. T. Hong, the MACRO Collaboration}.
\newblock Multiple muon measurements with macro.
\newblock {\em hep-ex/9410001}, 1995.

\bibitem{PhysRevD.52.3793}
{M. Ambrosio et al, the MACRO collaboration}.
\newblock Vertical muon intensity measured with macro at the gran sasso
  laboratory.
\newblock {\em Phys. Rev. D}, 52:3793--3802, Oct 1995.

\bibitem{max}
Giuseppe Battistoni, Annarita Margiotta, Silvia Muraro, and Maximiliano Sioli.
\newblock {FLUKA as a new high energy cosmic ray generator}.
\newblock {\em Nuclear Instruments and Methods in Physics Research Section A:
  Accelerators, Spectrometers, Detectors and Associated Equipment},
  626-627:S191--S192, 2011.

\bibitem{macro93}
{S. Ahlen, the MACRO Collaboration}.
\newblock {Muon Astronomy with the MACRO Detector}.
\newblock {\em The Astrophysical Journal}, 412:301--311, 1993.

\bibitem{cosmogenic_bx}
{Davide D'Angelo, for the Borexino collaboration}.
\newblock {Cosmogenic Backgrounds in Borexino}.
\newblock {\em to.be.published}, 2012.

\bibitem{selvi1}
M.~Selvi.
\newblock Measurement of muon-induced neutrons with lvd.
\newblock {\em presentation at Cosmogenic Activity and Background Workshop,
  LBNL}, April 2010.

\bibitem{selvi2}
M.~Selvi.
\newblock {Background estimation for Xe1T}.
\newblock {\em presentation at Cosmogenic Activity and Background Workshop,
  LBNL}, April 2010.

\bibitem{kamland}
S.~Abe et~al.
\newblock {Production of radioactive isotopes through cosmic muon spallation in
  KamLAND}.
\newblock {\em Phys. Rev. C}, 81, February 2010.

\bibitem{bx_muons}
{G. Bellini et al, the Borexino collaboration}.
\newblock {Muon and Cosmogenic Neutron Detection in Borexino}.
\newblock {\em Journal of Instrumentation}, 6, 2011.

\bibitem{bx_paper}
{G. Alimonti et al, the Borexino collaboration}.
\newblock {The Borexino detector at the Laboratori Nazionali del Gran Sasso}.
\newblock {\em Nuclear Instruments and Methods in Physics Research Section A:
  Accelerators, Spectrometers, Detectors and Associated Equipment}, 600, 2008.

\bibitem{russian}
A.S. Malgin and O.G. Ryazhskaya.
\newblock {Neutrons from Muons Underground}.
\newblock {\em Physics of Atomic Nuclei}, 71(10):1769­1781, 2008.
\newblock {Original Russian Text published in Yadernaya Fizika (2008, Vol. 71,
  No. 10, pp. 1800­1811)}.

\bibitem{bx11c}
{G. Bellini et al, the Borexino collaboration}.
\newblock {First evidence of pep solar neutrinos by direct detection in
  Borexino}.
\newblock {\em Phys. Rev. Lett.}, 108, 2012.

\bibitem{lvd05}
{N. Agafonova et al, the LVD collaboration}.
\newblock {The Measurement of the Total Specific Muon-Generated Neutron Yield
  Using LVD}.
\newblock {\em {29th International Cosmic Ray Conference}}, 9:239--242, 2005.

\bibitem{wang}
Y.~F. Wang, V.~Balic, G.~Gratta, A.~Fass\`o, S.~Roesler, and A.~Ferrari.
\newblock {Predicting neutron production from cosmic-ray muons}.
\newblock {\em Phys. Rev. D}, 64, 2001.

\bibitem{galbiati}
C.~Galbiati.
\newblock {\em {Data Taking and Analysis of the Counting Test Facility of
  Borexino}}.
\newblock PhD thesis, Universit\`a degli Studi di Milano, 1999.

\bibitem{hagner}
T.~Hagner et~al.
\newblock {Muon-induced production of radioactive isotopes in scintillation
  detectors}.
\newblock {\em Astroparticle Physics}, 14(1):33--47, 2000.

\bibitem{bxCTF}
{H. Back et al, the Borexino collaboration}.
\newblock {CNO and pep neutrino spectroscopy in Borexino: Measurement of the
  deep-underground production of cosmogenic 11 C in an organic liquid
  scintillator}.
\newblock {\em Phys. Rev. C}, 74, 2006.

\bibitem{lvdlat}
L.~Pandola, V.~Tomasello, and V.~Kudryavtsev.
\newblock {Neutron- and muon- induced background in underground physics
  experiments}.
\newblock {\em presentation at: ILIAS 4th Annual Meeting}, 2007.

\bibitem{ds50}
{Peter D. Meyers, Cristiano Galbiati, Frank P. Calaprice}.
\newblock {DarkSide-50: A Direct Search for Dark Matter with New Techniques for
  Reducing Background}.
\newblock {\em Proposal, submitted by Princeton to DOE}, April 2011.

\bibitem{ctf}
{G Alimont, et al, the Borexino collaboration}.
\newblock {A large-scale low-background liquid scintillation detector: the
  counting test facility at Gran Sasso}.
\newblock {\em NIM A}, 406:411–426, 1998.

\bibitem{tyvek}
{L. Wang et al}.
\newblock {Study of Tyvek reflectivity in water}.
\newblock {\em Chinese Physics C}, 36(7):628--632, 2012.

\bibitem{pureH2O}
{M. R. Querry, D. M. Wieliczka, D. J. Segelstein}.
\newblock {Handbook of Optical Constants of Solids II}.
\newblock pages 1059--1077, 1991.

\bibitem{McCarty}
Kevin~B. McCarty.
\newblock {\em {The Borexino Nylon Film and the Third Counting Test Facility}}.
\newblock PhD thesis, Princeton University, 2006.

\bibitem{lsv}
{Alex W., P. Mosteiro, B. Loer, and F. Calaprice}.
\newblock {A Highly Efficient Neutron Veto Using Boron‐Loaded Liquid
  Scintillator}.
\newblock {\em AIP Conf. Proc.}, 1338:44--48, 2010.

\bibitem{dsg2}
{Peter D. Meyers, Cristiano Galbiati, Frank P. Calaprice}.
\newblock {DarkSide-50: A Direct Search for Dark Matter with New Techniques for
  Reducing Background}.
\newblock {\em Proposal, submitted by Princeton to DOE}, April 2011.

\end{thebibliography}

\end{document}